\documentclass[twoside,fleqn]{article}
\usepackage{espcrc2}
\usepackage{amssymb,amsmath}

\usepackage{graphicx}

\newcommand{\half}{\frac{1}{2}}

\newcommand{\that}{\hat{t}}
\newcommand{\bra}{\langle}
\newcommand{\ket}{\rangle}
\newcommand{\Tr}{\operatorname{tr}}

\newcommand{\del}{\partial}
\newcommand{\om}{\omega}
\newcommand{\dl}{\delta}
\newcommand{\mhinv}{\,m_H^{-1}}

\newlength{\colw}
\newcommand{\cpaper}{\cite{Skullerud:2003ki}}

\title{Classical simulations and particle production in heavy-ion
  collisions} 

\author{Jon-Ivar Skullerud\address{ITF,
        University of Amsterdam, Valckenierstraat 65, 1018 XE
        Amsterdam, The Netherlands}}
\begin{document}
\makeatletter \@mathmargin = 0pt \makeatother

\begin{abstract}
The classical approximation may be applied to a number of problems in
non-equilibrium field theory.  The principles and limits of classical
real-time lattice simulations are presented, with particular emphasis
on the definition of particle numbers and energies and on applications
to the earliest stages of heavy-ion collisions.
\end{abstract}

\maketitle

\section{INTRODUCTION}

One of the most important predictions of lattice QCD is the existence
of a transition from ordinary hadronic matter, where quarks and gluons
are confined, to a quark--gluon plasma at high temperatures.  Current
estimates are that this transition takes place at $T_c\sim150-170$
MeV, and is most likely a crossover at zero chemical potential.  At
higher chemical potential a first-order phase transition is predicted,
ending in a tricritical point at $\mu^E_B\sim100-500$ MeV.

These predictions are currently being put to the test at RHIC and
other heavy-ion colliders.  However, interpreting the results of these
experiments and comparing them to lattice QCD predictions is far from
straightforward.  One important reason for this is that the process to
a large extent takes place out of thermal equilibrium, and equilibrium
field theory methods such as lattice Monte Carlo are therefore not
sufficient.

One major, still unsolved puzzle of heavy-ion physics is whether the
system ever actually reaches thermal equilibrium, and if so, what the
equilibration time is.  Many aspects of the collision can be
successfully described (eg, using hydrodynamics \cite{Kolb:2003dz}) by
assuming a very short equilibration time of 1 fm$/c$, but it is far from
understood how this would come about.  Clearly, a proper understanding
of the thermalisation process is essential if we are to have a
coherent and reliable description of the heavy-ion collision.

At late stages of the collision process, as the system expands, the
particles eventually decouple.  This typically occurs in two phases:
first, inelastic collisions cease, causing the ratios of different
particle species to be fixed (chemical freeze-out); later, also the
mean free path for elastic collisions becomes too large and the
momentum distribution of the particles deviates from that of thermal
equilibrium (thermal or kinetic freeze-out).  It is at this point that
the final particle yields of the collision are fixed.  Thus,
non-equilibrium dynamics is needed to understand also this aspect of
the process.

Another field where non-equilibrium field theory is needed, is
early-universe physics.  Examples of non-equilibrium processes which
require a field-theoretical approach include (p)reheating, electroweak
and QCD phase transitions in the early universe, and baryogenesis.

Field theories out of equilibrium is a notoriously difficult problem
to study nonperturbatively.  A large number of approaches have been
employed, including Hartree and large-N approximations
\cite{Cooper:1994hr,Boyanovsky:1995yk}, Dyson--Schwinger-related
approaches based on the 2PI effective action
\cite{Berges:2000ur,Berges:2003pc}, and kinetic theory
\cite{Baier:2000sb}.  All of these have their strengths as well as
drawbacks and limitations.  Within its area of applicability, the
classical approximation has the advantage of being fully
nonperturbative, easy and relatively inexpensive to implement
numerically, and straightforwardly applicable to gauge theories.  The
major drawback is obviously that quantum effects are not taken into
account.

\section{THE CLASSICAL APPROXIMATION}

Classical statistical physics is quantum statistical physics in the
limit of large occupation numbers.  It follows that the classical
approximation can be used when the occupation numbers of the relevant
or dominant modes of the system are large.  In the context of
heavy-ion collisions it is arguable that the multiplicity of soft
gluons in the initial (pre-equilibrium) stages is very high, and the
classical approximation may therefore be valid.

The classical approximation has also been applied to the chiral
dynamics during freeze-out.  In the linear sigma model and related
models, the effective potential of the chiral $(\sigma,\vec{\pi})$
field changes from symmetric to Mexican-hat type as the temperature
drops, giving rise to an instability where the low-momentum modes
increase exponentially.  In this scenario, the resulting high
occupation numbers justify the use of the classical approximation.

A particular hazard with simulating classical dynamics on a lattice is
connected with high-momentum lattice artefacts.  These will in general
interact with the soft modes which carry the interesting physics and
for which the classical approximation is in principle valid.  If there
is sufficient strength in the hard modes, they may equilibrate
classically with the soft modes on a much shorter timescale than that
of the interesting physics \cite{Moore:2001zf}.  In that case, not
only the hard modes, but the entire system will be dominated by
classical lattice artefacts.

To avoid this problem, it is important that the high-momentum modes
should be, and remain, strongly suppressed.  At early stages, this can
be ensured by choosing appropriate initial conditions.

The initial conditions are a crucial part of the simulation.  They
should reflect the salient features of the system at the outset.  This
is also the only place where information about the quantum nature of
the real world enters into the simulation.  A quantum system may be
represented as an ensemble of classical configurations initially
distributed according to quantum statistics \cite{Salle:2000hd}.  One
example of this may be to choose the 2-point correlators of the fields
and their canonical momenta to obey the Bose--Einstein distribution
for free fields at some temperature $T$, after subtracting the quantum
vacuum fluctuations.  For scalar fields $\phi$ with momentum fields
$\pi$ and mass $m$ one would then have
\begin{align}
\bra\phi(\vec{k},t=0)\phi(-\vec{k},0)\ket &= 
 \frac{1}{\om_k}\frac{1}{e^{\om_k/T}-1}\,, \\
\bra\pi(\vec{k},0)\pi(-\vec{k},0)\ket &= 
 \om_k\frac{1}{e^{\om_k/T}-1} \,,
\end{align}
where $\om_k=\sqrt{\vec{k}^2+m^2}$.
Such an initialisation also provides an exponential cutoff for the
hard modes, which will help in avoiding the dangerous lattice
artefacts.  Following this, each configuration evolves independently
according to the classical hamiltonian equations of motion, and
time-dependent correlators are computed as averages over the initial
conditions.

Given sufficient time, the system will eventually thermalise
classically, resulting in classical equipartition $n_k = \om_k/T$ and
giving rise to Rayleigh--Jeans type divergences.  The hope is that
this will happen on much longer time scales than those under
consideration in the simulation.

\section{DISTRIBUTION FUNCTIONS}

There is no unique definition of local particle numbers and energies
for interacting fields out of equilibrium.  Still, the system may
exhibit effective particle-like behaviour, which may be used to
characterise the approach to thermal equilibrium or to an
equilibrium-like distribution.  Given a definition of local particle
numbers, these can also be used to give an effective description of
the system in terms of kinetic theory.

The effective particle numbers may be extracted from the two-point
field correlators, which in the free-field case (where the particle
description is appropriate and well-defined) contain all the
information there is about the system.  For example, for a
homogeneous, free scalar field 
we have
\begin{align}
\bra\phi(\vec{k},t)\phi(-\vec{k},t)\ket
 & = \frac{1}{\om_k}\left(n_k+\half\right) \\
\bra\pi(\vec{k},t)\pi(-\vec{k},t)\ket
 & = \om_k\left(n_k+\half\right)
\end{align}
where $n_k$ is the occupation number for the mode with momentum
$\vec{k}$ and $\om_k$ is the associated energy.  For interacting
fields, this may in turn be used as a {\em definition} of the
instantaneous particle numbers and energies $n$ and $\om$
\cite{Aarts:1999zn,Salle:2000hd}:
\begin{align}
n_k(t)\! +\! \half & \equiv
 \sqrt{\bra\phi(\vec{k},t)\phi(-\vec{k},t)\ket 
  \bra\pi(\vec{k},t)\pi(-\vec{k},t)\ket}\,, \\
\om_k(t) & \equiv
 \sqrt{\frac{\bra\pi(\vec{k},t)\pi(-\vec{k},t)\ket}
        {\bra\phi(\vec{k},t)\phi(-\vec{k},t)\ket}} \, .
\end{align}
In the classical approximation the $1/2$ is left out.

In a non-abelian gauge theory, the correlation functions will in
general be gauge dependent, so the distribution functions will contain
ambiguities due to the gauge choice.  This ambiguity may be removed by
constructing gauge invariant correlators using parallel transporters;
however, this introduces path dependence.  In particular, in a lattice
regularisation there is in general no one preferred path between two
points.

Although the distribution functions are not unique, all physical
observables extracted from them, such as masses, temperatures and
chemical potentials, should not depend on the definition and in
particular on the gauge.  As long as these quantities are not
well-defined on the other hand (such as when the system is very far
from equilibrium and the quasiparticle picture does not apply), one
may expect ``masses'' and ``temperatures'' to be definition-dependent.
Thus, studying the gauge dependence (or path dependence) of
distribution functions may serve the double purpose of monitoring the
approach to equilibrium and verifying the validity of the approach
used.

One natural choice of gauge is the Coulomb gauge, which is a smooth
gauge.  In a system with spontaneously broken gauge symmetry (e.g., a
Higgs system), the unitary gauge, where the (effective) Higgs field
has only one non-zero, real component, is another natural choice.
Other gauges, such as maximal abelian gauge, axial gauges or random
gauge, may also be considered.

In the Coulomb gauge, the gauge potential $A_i$ (but not its conjugate
momentum $E_i$) is purely transverse, and it can be shown that the
transverse free correlators behave analogously to the scalar case
\cpaper.  Thus the particle numbers and
energies can be defined as
\begin{equation}
n_k \equiv \sqrt{D^A_T(k)D^E_T(k)}\,, \qquad
\om_k \equiv \sqrt{\frac{D^E_T(k)}{D^A_T(k)}} \, .
\end{equation}
Here $D^A_T, D^E_T$ are the transverse $A$- and $E$-cor\-re\-la\-tors
respectively, constructed from the two-point functions $\bra
\Phi^a_i(\vec{k})\Phi^b_j(-\vec{k})\ket =
\dl^{ab}C^{\Phi\Phi}_{ij}(\vec{k}); \Phi=A,E$ according to
\begin{align}
C^{AA}_{ij}(\vec{k}) &=
\Bigl(\dl_{ij}-\frac{k_ik_j}{k^2}\Bigr)D_T^A(k)\, ,\\
C^{EE}_{ij}(\vec{k}) &=
\Bigl(\dl_{ij}\!-\!\frac{k_ik_j}{k^2}\Bigr)D_T^E(k)
 +\frac{k_ik_j}{k^2}D_L^E(k)\,.
\end{align}

\begin{figure}
\includegraphics[width=\colw,clip]{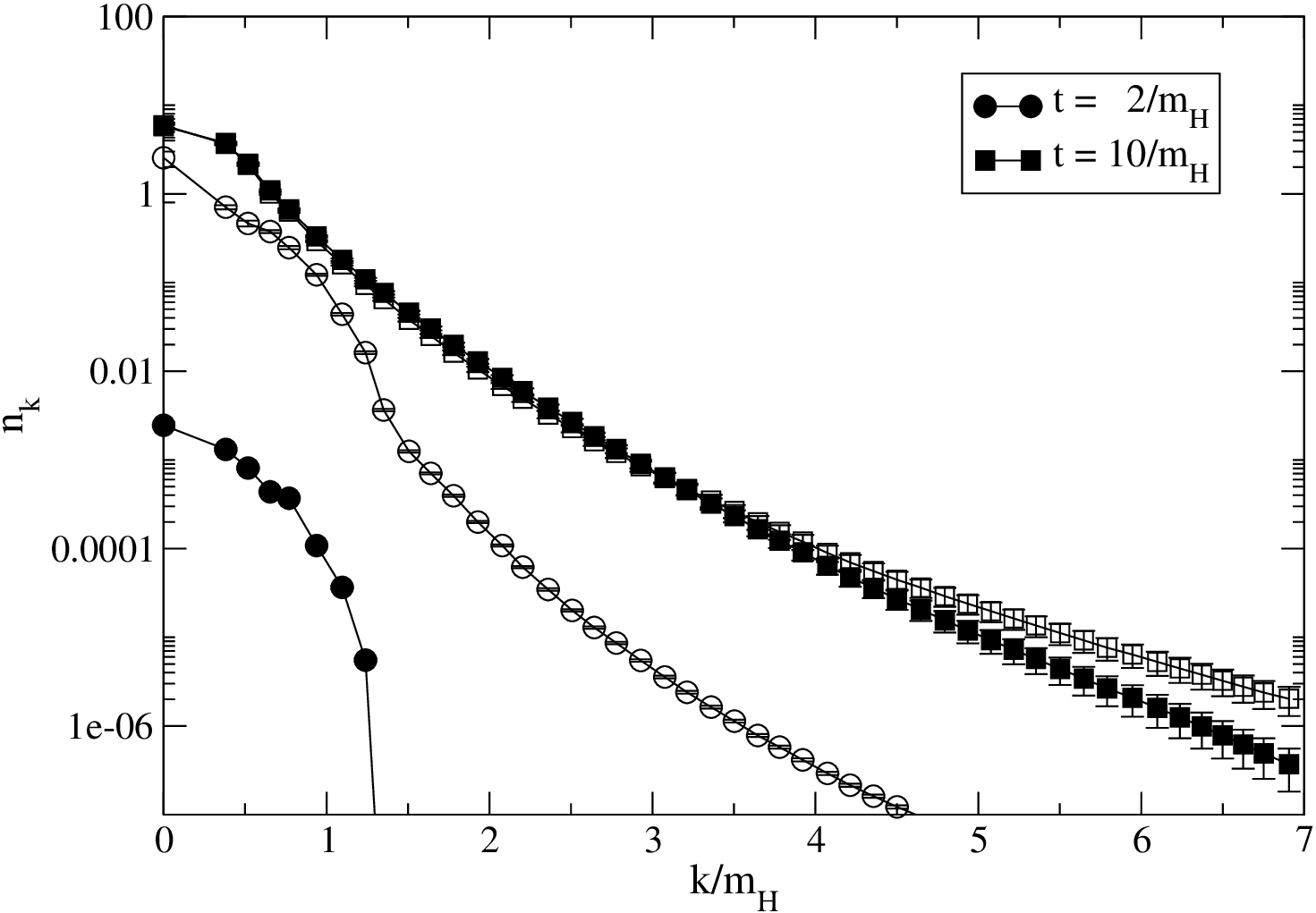}\\
\includegraphics[width=\colw,clip]{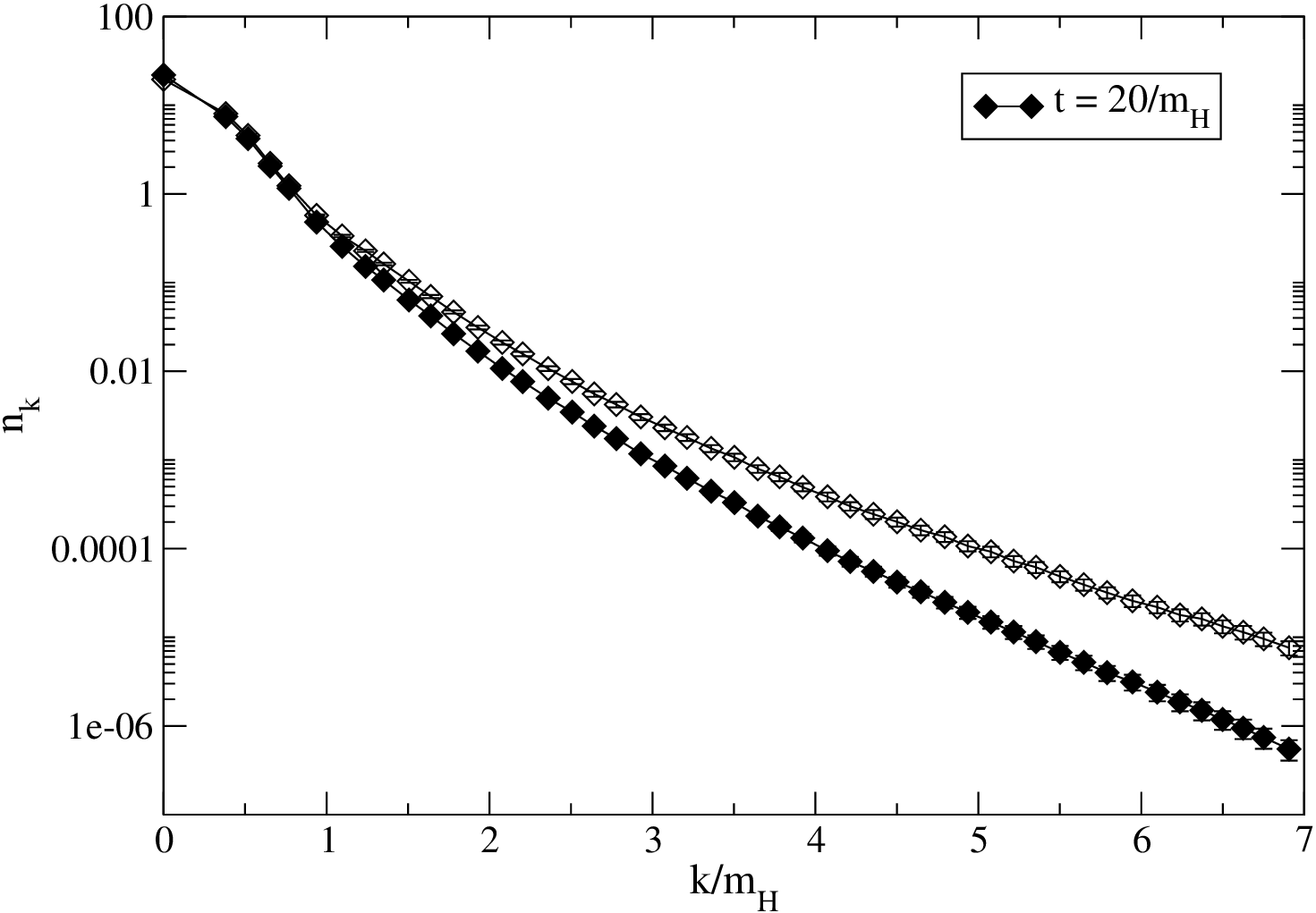}\\
\includegraphics[width=\colw,clip]{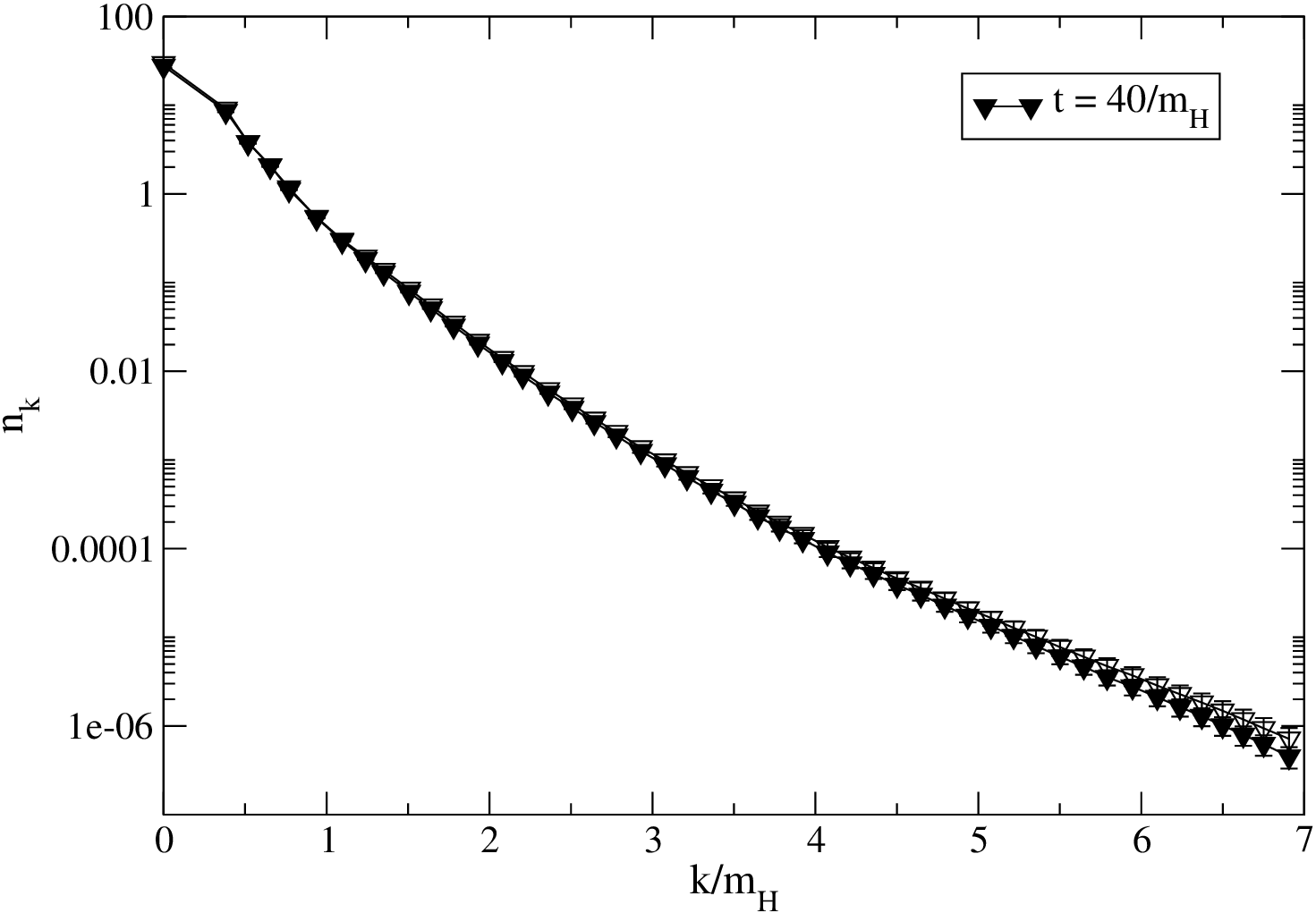}
\caption{Gauge dependence of gauge particle numbers in the
  SU(2)--Higgs model \protect\cpaper\ at different times.  The filled
  symbols are particle numbers obtained in the Coulomb gauge, while
  the open symbols are for the transverse modes in the unitary gauge.}
\label{fig:n-gauge}
\end{figure}
In figure \ref{fig:n-gauge} the gauge dependence of effective particle
numbers is illustrated in the SU(2)--Higgs model \cpaper.  In this
case, the system was prepared in such a way that all the energy
initially was in the Higgs field, while the gauge potential was
initialised to zero.  Since the angular modes of the Higgs fields are
absorbed into the gauge fields in the unitary gauge, the inital
occupation numbers are very different.  However, already after
$t=10\,\mhinv$ the two distributions appear almost identical.  However,
while the particle numbers in the Coulomb gauge change very little
from here on, in the unitary gauge they continue to fluctuate and it
is only from $t\approx40\,\mhinv$ on that one with some confidence can
claim the numbers are gauge independent.  This agrees roughly with the
point where the dispersion relation in the unitary gauge begins to
show stable, particle-like behaviour in this particular model.

In a non-homogeneous system, and in general in a kinetic-theory
description, it is appropriate to think in terms of {\em local}
particle numbers $n(\vec{x},\vec{k},t)$.  These may be related to the
Wigner functions constructed from gauge invariant two-point functions
\cite{Aarts:1999zn}, or more generally to the two-point functions
fourier transformed on a region $R(\vec{x})$ centred on $\vec{x}$.  In
the case of a scalar theory we may have (suppressing the common
$t$-coordinate for brevity)
\begin{align}
C_{\phi\phi}(\vec{x},\vec{k}) &=
 \frac{1}{\Omega_R}\int_{R(\vec{x})}\!\!\!\!\!\!\!d^3z\,d^3y\,
  e^{-i\vec{k}\cdot(\vec{y}-\vec{z})} \bra\phi(\vec{y})\phi(\vec{z})\ket \,,\\
C_{\pi\pi}(\vec{x},\vec{k}) &=
 \frac{1}{\Omega_R}\int_{R(\vec{x})}\!\!\!\!\!\!\!d^3z\,d^3y\,
  e^{-i\vec{k}\cdot(\vec{y}-\vec{z})}\bra\pi(\vec{y})\pi(\vec{z})\ket \,.
\end{align}
Here $\Omega_R$ is the volume of the region $R(\vec{x})$.  This
coarse-graining creates an intrinsic unsharpness in the momentum
$\vec{k}$ and position $\vec{x}$ of the quasiparticles, given by the
size (and shape) of the region $R$.

If the system under consideration is homogeneous, we may improve
statistics by performing an average over all space.  This can be shown
to be equivalent to local averaging in momentum space, with a weight
function $w$ depending on the size and shape of $R$:
\begin{equation}
\begin{split}
C(\vec{k},t) &= \frac{1}{V}\int\!\! d^3\!x\,C(\vec{x},\vec{k},t) \\
 &= \frac{1}{V}\int\!\! d^3\!x\, C_{R(\vec{x})}(\vec{k},t) \\
 &= \sum_{\vec{p}}w(\vec{p}-\vec{k})C_V(\vec{p},t) \, ,
\end{split}
\end{equation}
where $C_V$ denotes the correlation function evaluated on the total
volume $V$, and the sum is over the discrete momenta available on this
volume.  In practice, it is simpler to work backwards, choosing a
simple form of momentum-space averaging which may correspond to
rather complicated spatial regions.  For instance, binning in the
absolute value of the momentum,
\begin{equation}
w(\vec{p}-\vec{k})
 \propto \Theta(|\vec{k}|-|\vec{p}|+\frac{\Delta}{2})
 -\Theta(|\vec{k}|-|\vec{p}|-\frac{\Delta}{2}) \,,
\end{equation}
corresponds to spherical shells with thickness approximately
$1/\Delta$ in position space.

\section{PURE YANG--MILLS}

At the earliest stages of heavy-ion collisions, the gluon density is
expected to be so high that the classical approximation can be
justified.  The same approximation also justifies ignoring the
back-reaction of the quarks, since their number density will be much
lower; leaving us with classical Yang--Mills equations of motion,
which may be solved numerically on a lattice.

The lattice equations of motion in the temporal gauge ($A_0=0$) read
\begin{equation}
\del_tE^a_i(x) = D^{ab}_j\Tr\bigl[it^bU_{ji}(x)\bigr]
\end{equation}
where
\begin{equation}
E^a_i(x) = F^a_{0i}(x)
 = \Tr\bigl[t^aU_i(x)U^\dagger_i(x+\that)\bigr]
\end{equation}
is the canonical momentum to $A^a_i$.  Here, $\del_t$ denotes the
backward lattice derivative, while $D_j$ is the backward covariant
lattice derivative.  The equations of motion for $A^a_0$ constitute
the Gauss constraint,
\begin{equation}
D^{ab}_i E^b_i = 0 \, ,
\end{equation}
which must be satisfied by the initial conditions but is conserved by
the equations of motion.

The initial gluon fields should be related to the gluon distributions
of the two colliding nuclei: in principle they should just be the
superposition of two Lorentz-boosted nuclear gluon distributions.
Simulations have been carried out over a number of years by Krasnitz,
Nara and Venugopalan \cite{Krasnitz:1999wc,Krasnitz:2002mn} (see also
\cite{Lappi:2003bi}) using the ``colour glass condensate'' model of
the nuclear wave function to provide the initial conditions.  In these
studies, the numerical work has been simplified by considering only
the mid-rapidity region where the physics is assumed to be
boost-invariant.  This reduces the system to effectively 2+1
dimensions.  With these assumptions, the authors have been able to
provide an estimate of the initial energy density and gluon
distribution which may be used as input into hydrodynamic or kinetic
calculations.

An alternative approach would be to determine the nuclear gluon field
from e.g.\ a bag model, give this a Lorentz boost, and perform a
2+1+1-dimensional simulation with the longitudinal lattice spacing
$a_z=a_\perp/\gamma$, where $a_\perp$ is the lattice spacing in the
transverse ($x,y$) direction.  Work is underway to implement this.

\section{SUMMARY}

The classical approximation may be applied to a range of problems in
non-equilibrium field theory where occupation numbers are high, such
as the earliest stages of heavy-ion collisions.  It has the advantage
of being non-perturbative and computationally relatively inexpensive.
Effective particle numbers may be defined out of equilibrium in a
self-consistent manner, and their gauge dependence (or that of derived
quantities such as masses and temperatures) can be used as a check on
the validity of the quasi-particle picture.

\section*{Acknowledgments}

This work was supported by FOM/NWO.  I am thankful to Jan Smit and
Anders Tranberg for numerous fruitful discussions.


\end{document}